\documentclass[12pt, onecolumn]{article}
\usepackage{adjustbox}
\usepackage[linesnumbered, ruled]{algorithm2e}
\usepackage{latexsym, indentfirst}
\usepackage{amssymb,amsmath,amsthm,float}
\usepackage{mathrsfs}
\usepackage{extarrows}
\usepackage{color,soul}
\usepackage{caption}
\usepackage{makecell}
\usepackage{longtable}
\usepackage{dcolumn}
\usepackage{array}
\usepackage{booktabs}
\usepackage{subfigure}
\usepackage{authblk} 
\usepackage{multirow}
\usepackage{rotating}
\usepackage{galois} 
\usepackage{graphicx}
\usepackage{subfigure}
\usepackage{enumitem}
\usepackage{epstopdf}
\usepackage{bm}
\usepackage{titlesec}
\usepackage{MnSymbol}
\usepackage{colortbl} 
\usepackage{xcolor}
\usepackage{threeparttable}
\usepackage[a4paper, 
left=2.54cm,    
right=2.54cm,   
top=2.54cm,     
bottom=2.54cm,  
footskip=1.5cm]{geometry}
\usepackage{setspace}
\allowdisplaybreaks

\numberwithin{equation}{section}

\usepackage{hyperref}    
\hypersetup{hypertex=true,
colorlinks=true,
linkcolor=blue,
anchorcolor=blue,
citecolor=blue}

\renewcommand{\abstract}{{\noindent\bf Abstract.}}

\newtheorem{theorem}{Theorem}

\newtheorem{lemma}{Lemma}
\newtheorem{remark}{Remark}
\newtheorem{corollary}{Corollary}

\def\b{\beta}                                   \def\a{\alpha}
                                  \def\g{\gamma}
\def\p{\phi}                                    \def\t{\theta}
\def\l{\lambda}                                 
                                 
                                  \def\e{\varepsilon}

\def\k{{(k)}}                                   \def\mm{{(m)}}

\def\KL{\mathrm{KL}}                            
                              
\def\Pr{\mathrm{Pr}}

\begin{document}

\title{\textbf{\large Semiparametric model averaging for high-dimensional quantile regression with nonignorable nonresponse}}
\author[1]{Wei Xiong}
\author[2]{Dianliang Deng}
\author[1]{Dehui Wang\thanks{The corresponding address: wangdehui@lnu.edu.cn}}
\affil[1]{\footnotesize School of Mathematics and Statistics, Liaoning University, Shenyang, China}
\affil[2]{Department of Mathematics and Statistics, University of Regina, SK, Canada}
\date{}
\maketitle
\begin{abstract}
	Model averaging has demonstrated superior performance for ensemble forecasting in high-dimensional framework, its extension to incomplete datasets remains a critical but underexplored challenge. Moreover,  identifying the parsimonious model through averaging procedure in quantile regression demands urgent methodological innovation. In this paper, we propose a novel model averaging  method for high-dimensional quantile regression with nonignorable missingness. The idea is to relax the parametric constraint on the conditional distribution of respondents, which is constructed through the two-phase scheme: (i) a semiparametric likelihood-based estimation for the missing mechanism, and (ii) a semiparametric weighting procedure to combine candidate models. One of pivotal advantages is our SMA estimator can asymptotically concentrate on the optimally correct model when the candidate set involves at least one correct model. Theoretical results show that the estimator achieves asymptotic optimality even under complex missingness conditions. Empirical conclusions illustrate the efficiency of the method.
\end{abstract}

\section{Introduction}

Statistical inference of high-dimensional data has become increasingly important across many scientific fields. In the past two decades, there has been a number of contributive literature such as \hyperlink{ref:Fan2008}{Fan and Lv (2008)}, \hyperlink{ref:Wang2011}{Wang (2011)}, \hyperlink{ref:Li2012}{Li et al. (2012)}. Compared to least squares model, quantile regression (\hyperlink{ref:Koenker2005}{Koenker 2005}) offers a robust alternative to check the effect of high-dimensional covariates on the entire response distribution. To efficiently select active predictors, \hyperlink{ref:He2013}{He et al. (2013)} introduced a criterion of variable screening by marginal quantile utility. More recently, \hyperlink{ref:Kong2019}{Kong et al. (2019)} and \hyperlink{ref:Jiang2024}{Jiang et al. (2024)} modified the method and used it to refine variable selection in ultra-high dimensional quantile model, respectively.

One of central challenges in high-dimensional regression analysis is prediction. Model averaging emerges as a compelling alternative to classical selection procedures (such as AIC and BIC), effectively addressing the inherent uncertainty associated with selecting a single model. Following a series of pioneering contributions by \hyperlink{ref:Yang2001}{Yang (2001)}, \hyperlink{ref:Hjort2003}{Hjort and Claeskens (2003)}, and \hyperlink{ref:Hansen2007}{Hansen (2007)}, theoretical development of frequentist model averaging has been rapidly advancing. For quantile regression, \hyperlink{ref:Shan2009}{Shan and Yang (2009)} applied the adaptive aggregating algorithm for mixing different specifications, \hyperlink{ref:Lu2015}{Lu and Su (2015)} proposed the jackknife model averaging (JMA) procedure for the divergent number of predictors, \hyperlink{ref:Wang2023}{Wang et al. (2023)} combined JMA with variable screening for the ultra-high dimensional case, \hyperlink{ref:Xu2023}{Xu et al. (2023)} studied the application of cross-validation model averaging (CVMA) in functional linear model, and among others.

Abovementioned methodologies are established with fully available dataset. In practice, nonresponse is frequently encountered during data collecting. Naive exclusion of missingness may causes biased results when the response mechanism depends on unobserved parts --- the scenario defined as missing not at random (MNAR) (\hyperlink{ref:Little2002}{Little and Rubin 2002}; \hyperlink{ref:Kim2021}{Kim and Shao 2021}). Statistical inference of MNAR data has been greatly developed, including dimension reduction and estimation (e.g., \hyperlink{ref:Kim2011}{Kim and Yu 2011}, \hyperlink{ref:Miao2016}{Miao and Tchetgen 2016},  \hyperlink{ref:Shao2016}{Shao and Wang 2016}, \hyperlink{ref:Morikawa2021}{Morikawa and Kim 2021}, \hyperlink{ref:Zhang2020CSDA}{Zhang et al. 2020}, \hyperlink{ref:Zhao2021}{Zhao et al. 2021}). In recent years, researchers begin to take focus on model averaging with missing data, and most of studies are regarded with ignorable mechanisms. For example, \hyperlink{ref:Wei2021}{Wei et al. (2021)}, \hyperlink{ref:Xie2021}{Xie et al. (2021)} used frequentist model averaging and inverse probability weighting (IPW) to predict the conditional mean function, respectively. However, for MNAR, the literature currently lacks model averaging procedures for either the response mechanism or the conditional quantile function (CQF). We wish to address the void in this paper.

The objective of the paper is to contribute a novel semiparametric model averaging (SMA) procedure for predicting CQF. To the best of our knowledge, most of methodologies can hardly skirt the estimation of MNAR mechanism. As pointed out by \hyperlink{ref:Li2023}{Li et al. (2023)}, instability will be occurred through the moment estimating equation. They suggested a likelihood estimation under the parametric assumption of the conditional distribution of respondents. In the high-dimensional case, we derive a weighted estimator that combines semiparametric likelihood and model averaging. The advantage is to avoid potential misspecification of the conditional distribution, and the computational burden from nonparametric parts can be practically reduced by a sampling technique. We verify that the estimator of nuisance parameter is consistent when the correct model are contained in the candidate set. 

Asymptotic theory of the weight estimator is another interesting topic in model averaging literature, while related studies are almost focused on least squares regression (\hyperlink{ref:Zhang2019}{Zhang and Liu 2019}; \hyperlink{ref:Zhang2020}{Zhang et al. 2020}; \hyperlink{ref:Fang2023}{Fang et al. 2023}). Recent advance in \hyperlink{ref:Xu2023}{Xu et al. (2023)} addressed the gap by establishing the over-consistency (the concept refers to \hyperlink{ref:Chen2023}{Chen et al. 2023}) of weights of CVMA, while the behavior of the weight assigned to the optimally correct model remains to be developed. The current paper enriches the theoretical results of proposed procedure in two scenarios: (i) when the candidate set contains at least one correct model, SMA can asymptotically put the weight one to the optimally correct model, i.e., parsimony; (ii) when all candidates are misspecified, SMA preserves the same asymptotic optimality derived from full data in the sense that its prediction risk via quantile loss is asymptotically identified to the best--but--infeasible averaging estimator. Note that above conclusions are investigated on the continuous weight support with ``nonnegative and sum-up-to-one'', it is regarded as a supplement beyond conventional least squares framework.

Numerically we propose a sampling approximation for SMA procedure to alleviate computational burden induced by nonparametric regression. Besides, to address the curse of dimensionality and efficiently construct candidate sets, we also develop a greedy post-selection algorithm based on grouped variable importance. We empirically compare the proposed estimation to existing methods of model selection and model averaging. Numerical results demonstrate that SMA can effectively identify the optimal candidate and has satisfactory predictive performance. Further, our method is more robust than others to against misspecification of the response mechanism.

The reminder of this paper is organized as follows. Section 2 describes the model framework, the SMA procedure for MNAR mechanism, and investigates the consistency of nuisance parameter. Section 3 proposes the SMA criterion to predict CQF. Section 4 establishes the theoretical properties of the CQF estimator. Section 5 introduces an efficient computational sampling method to evaluate conditional expectations of respondents. Numerical studies are presented in Section 6. A post-selection algorithm for preparing candidate sets is described in Appendix. Supplementary simulation results and technical details of proofs are attached in online supplementary material, respectively.

\section{Model and estimation for MNAR mechanism}

\subsection{Model specification}

Let $\mathcal{D}_n=\{(\bm{x}^\top_i,y_i)^\top:i=1,...,n\}$ be the set of independent and identically distributed (i.i.d.) copies of $(\bm{x}^\top,y)^\top$, where $\bm{x}$ is a $p$-dimensional covariate vector that can be fully observed, and $y$ is the outcome with some of observations are nonresponse.
Given a probability level $\tau\in(0,1)$, the CQF of the response takes the form as
\begin{equation}
Q_\tau(y_i|\bm{x}_i)=\bm{\b}^\top(\tau)\bm{x}_i\triangleeq\mu_i(\tau),
\label{sec2.eq1}
\end{equation} 
where $\bm{\b}(\tau)$ is the parameter vector that may depends on $\tau$. Obviously (\ref{sec2.eq1}) is equivalent to the following quantile regressive model:
$$
y_i=\mu_i(\tau)+\e_i(\tau),
$$
where $\e_i(\tau)$'s are i.i.d. random errors such that $\Pr(\e_i(\tau)\leq 0|\bm{x}_i)=\tau$. Hereafter we omit the notation ``$(\tau)$'' for simplicity. Since some of $y_i$'s are subject to missingness, we define $r_i$ as the binary response indicator such that $y_i$ is observed when $r_i=1$ and $y_i$ is missing when $r_i=0$. The response mechanism is assumed as a logistic model that
\begin{equation}
\Pr(r_i=1|\bm{x}_i,y_i) = \frac{1}{1+\exp\left(\varphi_0+\sum^p_{j=1}\varphi_jx_{ij}+\g y_i\right)}
\triangleeq \pi_i(\bm{\t}),
\label{sec2.eq2}
\end{equation}
where $\bm{\t}=(\varphi_0,...,\varphi_p,\g)^\top$ is the unknown parameter vector. 

Through the paper, $p$ is allowed to be divergent with respect to $n$, and the main interest is to effectively predict (\ref{sec2.eq1}) by the pairs $(\bm{x}^\top_i,y_i)^\top$ among $r_i=1$. Due to MNAR, we firstly need to evaluate the nuisance parameter in (\ref{sec2.eq2}). 

\subsection{Semiparametric model averaging for response mechanism}

As we can see, evaluating $\bm{\t}$ is challenging because (i) the nonresponse variable is involved in the mechanism model and (ii) the distribution of $y|\bm{x},r=1$ may be inconsistent with that of $y|\bm{x}$, one cannot directly construct the likelihood function for (\ref{sec2.eq2}). 
Motivated by \hyperlink{ref:Kim2021}{Kim and Shao (2021)}, we can find
$$
\frac{1-\Pr(r_i=1|\bm{x}_i;\bm{\t})}{\Pr(r_i=1|\bm{x}_i;\bm{\t})} =
E\left(\frac{1-\pi_i(\bm{\t})}{\pi_i(\bm{\t})}\bigg|\bm{x}_i,r_i=1\right) =\exp\left\{\varphi_0+\sum^p_{j=1}\varphi_jx_{ij}+m_1(\bm{x}_i;\g)\right\},
$$
where $m_1(\bm{x}_i;\g)=\log E(\exp(\g y)|\bm{x}_i,r_i=1)$ replaces the nonresponse part. To avoid the potential misspecification, we allow a nonparametric form on $y|\bm{x},r=1$, hence the question is transformed into estimating parameter in $\Pr(r_i=1|\bm{x}_i;\bm{\t})$ by a semiparametric method.  

To describe the SMA procedure, we first use nonparametric regression for $m_1(\bm{x}_i;\g)$. Note that
$$
m_1(\bm{x};\g)
=\log E\{r\exp(\g y)|\bm{x}\} - \log\Pr(r=1).
$$
For a kernel function $K_h(\cdot)=K(\cdot/h)$ and an approximated bandwidth $h=h_n$, the corresponding estimator is 
\begin{equation}
\hat{m}_1(\bm{x}_i;\g) = \log\frac{\sum^n_{j=1}r_j\exp(\g y_j)K_h(\bm{x}_i-\bm{x}_j)}{\sum^n_{j=1}K_h(\bm{x}_i-\bm{x}_j)}
- \log\frac{\sum^n_{j=1}r_j}{n}.
\label{sec2.eq3}
\end{equation}
Having parametrized $\Pr(r_i=1|\bm{x}_i;\bm{\t})$, we can decompose it into multiple candidates and aggregate them through model averaging.

Let $\bm{\t}^\k=(\bm{\varphi}^{\k^\top},\g)^\top$, and $\mathcal{S}_1=\{\Pr^\k(r=1|\bm{x};\bm{\t}^\k):k=1,...,S_1\}$ be the candidate model set of $\Pr(r=1|\bm{x};\bm{\t})$. Note that the estimation of $\bm{\t}$ will participate in predicting CQF, it imposes the requirement on $\mathcal{S}_1$ that: there exists at least one candidate model that sufficiently covers active predictors, i.e., it is preferable to contain correct models. Of this view we make a nested assumption on $\mathcal{S}_1$ for both theoretical and practical convenience. Suppose that all linear predictors are decreasingly ranked by their importance, the $k$-th candidate is formed as
$$
-\mathrm{logit}\left\{\Pr^\k(r=1|\bm{x};\bm{\t}^\k)\right\} 
=
\bm{\varphi}^{\k^\top}\bm{x}^\k+m_1(\bm{x};\g)
= 
\sum^{|\mathcal{M}_k|}_{j=1}\varphi_jx_j+m_1(\bm{x};\g),
$$
where $\mathcal{M}_1\subset...\subset\mathcal{M}_{S_1}\subset\{1,...,p\}$ are index sets of linear predictors and $|\mathcal{M}_k|$ is the cardinality.
The parameter can be estimated as
$$
\hat{\bm{\t}}^\k_n=\left(\hat{\bm{\varphi}}^{\k^\top}_n,\hat{\g}^\k_n\right)^\top=
\mathop{\arg\min}_{\bm{\t}^\k\in\Phi_k\times\Gamma}L^\k_n\left(\bm{\t}^\k\right),
$$
where
$$
L^\k_n(\bm{\t}^\k) = n^{-1}\sum^n_{i=1}\left[(r_i-1)\left(\bm{\varphi}^{\k^\top}\bm{x}^\k_i+\hat{m}_1(\bm{x}_i;\g)\right)+\log\left\{1+\exp\left(\bm{\varphi}^{\k^\top}\bm{x}^\k_i+\hat{m}_1(\bm{x}_i;\g)\right)\right\}\right]
$$
is the negative log-likelihood function of the $k$-th model and $\hat{m}_1(\bm{x}_i;\g)$ is given in (\ref{sec2.eq3}). 

The remainder is to match weights over candidates. Let $\bm{\omega}=(\omega_1,...,\omega_{S_1})^\top\in\mathcal{H}_1$ be a vector where the $k$-th component $\omega_k$ represents the weight of $\hat{\bm{\theta}}^{(k)}_n$, and
$\mathcal{H}_1=\{\bm{\omega}\in[0,1]^{S_1}:\sum^{S_1}_{k=1}\omega_k=1\}$. Define $\hat{\g}(\bm{\omega})=\sum^{S_1}_{k=1}\omega_k\hat{\g}^\k_n$ and
$$
\hat{\eta}_i(\bm{\omega}) = \sum^{S_1}_{k=1}\omega_k\hat{\bm{\varphi}}^{\k^\top}_n\bm{x}^\k_i
+\hat{m}_1\left(\bm{x}_i;\hat{\g}(\bm{\omega})\right),
$$ 
the weighting criterion of SMA is proposed as
$$
\mathrm{SL}(\hat{\eta}(\bm{\omega}))
= \sum^{n}_{i=1}
\left[ (1-r_i)\hat{\eta}_i(\bm{\omega})-\log\left\{1+\exp\left(\hat{\eta}_i(\bm{\omega})\right)\right\}\right]
-\l_n\sum^{S_1}_{k=1}\omega_k|\mathcal{M}_k|,
$$
where $\l_n$ is a penalty scale depending on $n$. The regularized term is imposed to shrink the weight corresponding to overfitted models. Consequently, the SMA estimator of $\bm{\t}$ is
\begin{equation}
\hat{\bm{\t}}(\hat{\bm{\omega}}) =
\left(\hat{\bm{\varphi}}^\top(\hat{\bm{\omega}}),\hat{\g}(\hat{\bm{\omega}})\right)^\top
= \sum^{S_1}_{k=1} \hat{\omega}_k
\left(\hat{\bm{\varphi}}^{\k^\top}_n,\bm{0}^\top_{S_1-k},\hat{\g}^\k_n\right)^\top,
\label{sec2.eq4}
\end{equation}
where $\bm{0}_{S_1-k}$ is a $(|\mathcal{M}_{S_1}|-|\mathcal{M}_k|)$-dimensional vector of $0$, and 
$\hat{\bm{\omega}}= \mathop{\arg\max}_{\bm{w}\in\mathcal{H}_1}\mathrm{SL}(\hat{\eta}(\bm{\omega}))$.

\subsection{Asymptotic properties of the estimator}

This section devotes to study the theoretical performance of (\ref{sec2.eq4}). Under the linear specification, it is apparent to conform the identification of MNAR mechanism through distinguishing instrumental variable (\hyperlink{ref:Morikawa2021}{Morikawa and Kim 2021}, \hyperlink{ref:Li2023}{Li et al. 2023}). In the following discussion we default the estimator is identified.
Another fundamental permission is that both $S_1$ and $|\mathcal{M}_k|$'s can diverge with $n$ due to the high-dimensional setting (e.g., $|\mathcal{M}_{S_1}|$ is allowed to be upper-bounded by $O(p)$). 
Since (\ref{sec2.eq2}) performs as an agent of quantile regression, we particularly pay attention to the convergency of nuisance parameter. Denote $\bm{\t}_0$ as the ``true value'' of $\bm{\t}$, and
$$
\KL^\k(\bm{\t}) = E\left\{r\log\frac{\Pr(r=1|\bm{x};\bm{\t}_0)}{\Pr^\k(r=1|\bm{x};\bm{\t}^\k)}+(1-r)\log\frac{\Pr(r=0|\bm{x};\bm{\t}_0)}{\Pr^\k(r=0|\bm{x};\bm{\t}^\k)}\bigg|\bm{x}\right\}
$$ 
as the Kullback–Leibler (KL) divergence between $\Pr^\k(r=1|\bm{x};\bm{\t}^\k)$ and $\Pr(r=1|\bm{x};\bm{\t}_0)$. Commonly conditions (see in, \hyperlink{ref:White1982}{White 1982}) imply the existence of a unique ``pseudo-true'' $\bm{\t}^\k_*$, which satisfies
$$
\bm{\t}^\k_* = \mathop{\arg\min}_{\bm{\t}^\k\in\Phi_k\times\Gamma}E\left(\KL^\k(\bm{\t})\right).
$$
To prevent confusion we state some notations: (i) $\|\cdot\|$ denotes the Euclidean norm (i.e., for a matrix $A=(a_{ij})$, $\|A\|=(\sum_{i,j}a_{ij})^{1/2}$). (ii) $\mathcal{X}$ denotes the support of $\bm{x}$. (iii) $a\lesssim(\gtrsim) b$ denotes $a\leq(\geq) C\cdot b$ with a positive constant $C$. (iv) $f_{r}(\cdot)$ and $E_{r}(\cdot)$ denote conditional density function and conditional expectation among $r\in\{0,1\}$, respectively.
Firstly, we investigate consistency of candidate estimators, which requires the following regularity conditions: 
\begin{enumerate}
\item[(A1)]~$\Phi_k$'s and $\Gamma$ are compact over $k=1,...,S_1$.	
\item[(A2)]~$f_1(\bm{x})$ is bounded away from $0$ and $\infty$ uniformly over $\bm{x}\in\mathcal{X}$.
\item[(A3)]~The kernel $K(u)$ has a compact support and bounded derivatives until order $k$ ($k\geq 2$), $\int K(u)\mathrm{d}u=1$ and has zero moments of order up to $m-1$ and nonzero $m$-th order moment.	
\item[(A4)]~There exists $v\geq 4$ such that $E(y^v)$ is finite. On the other hand, $E\{\exp(v\g y)\}$ and $E_1\{\exp(v\g y)|\bm{x}\}f_1(\bm{x})$ are bounded away from $0$ and infinity on $\mathcal{X}$.
\item[(A5)]~$0<\pi_i(\bm{\t})<1$ uniformly over $\bm{\t}\in\Phi\times\Gamma$.
\item[(A6)]~As $n\to\infty$, the bandwidth satisfies: (i) $h\to0$, (ii) $n^{1-2/v}h^p/\log(n)\to\infty$, (iii) $\sqrt{n}h^{p+2k}/\log(n)\to\infty$ and (iv) $\sqrt{n}h^{2m}\to0$.
\item[(A7)]~$\max_{1\leq j\leq p}Ex^4_j=O(1)$.	

\end{enumerate}

Conditions (A1)--(A6) are widely employed in the literature of semiparametric inference for missing data (\hyperlink{ref:Kim2011}{Kim and Yu 2011}; \hyperlink{ref:Shao2016}{Shao and Wang 2016}; \hyperlink{ref:Morikawa2021}{Morikawa and Kim 2021}). In addition to establish the convergence of nonparametric estimator, they also imply that $Em_1(\bm{x};\g)$ has finite derivatives to order $2$. Condition (A7) is a mild restriction of the moment in high-dimensional data analysis (\hyperlink{ref:Wang2023}{Wang et al. 2023}), and is alternative for deriving the asymptotic properties of SMA estimator. 

\begin{lemma}
	Under conditions {\em (A1)--(A7)}, if $\max_{1\leq k\leq S_1}|\mathcal{M}_k|\lesssim n^{1/2}$, we have for $k=1,...,S_1$,
	$$
	\hat{\bm{\t}}^\k_n\stackrel{p}{\to}\bm{\t}^\k_*~~~\text{as}~~~n\to\infty.
	$$
\end{lemma}

From Lemma 1 it is easy to see that the estimator of a correct specification (in $\mathcal{S}_1$) is consistent to $(\bm{\t}^\top_0,0,...,0)^\top$. This result will play as a crucial premise to establish the consistency of SMA estimator. To the best of our knowledge, a well-working model averaging procedure should perform the largest propensity to the ``optimally correct'' (abbreviated OC) candidate.
Let $\mathcal{M}_T$ be the index set of significant linear predictors in $\Pr(r=1|\bm{x};\bm{\t})$, the $k$-th nested candidate is OC if $\mathcal{M}_T\subset\mathcal{M}_k$ and $\mathcal{M}_{k-1}\bigcap\mathcal{M}_T\subsetneq\mathcal{M}_T$ (i.e., the first $k-1$ candidates are underfitted and the last $S_1-k$ candidates are overfitted). We remain to make structural regularities on the candidate set to guarantee the reliability of subsequent inference.

We consider the scenario that $\mathcal{S}_1$ contains the OC model. Without loss of generality, the index of the OC candidate is specified as $1\leq s_*\leq S_1$. 
Similar to model averaging literature for generalized linear model, we define the infeasible KL loss among candidates as $\xi_n =\inf_{\bm{\omega}\in\mathcal{H}_1}\KL(\eta_*(\bm{\omega}))$, where
$$
\KL(\eta_*(\bm{\omega}))= \sum^n_{i=1}\left\{E(r|\bm{x}_i)\log\frac{\Pr(r=1|\bm{x}_i;\bm{\t}_0)}{\Pr(r=1|\bm{x}_i;\bm{\t}_*(\bm{\omega}))}+E(1-r|\bm{x}_i)\log\frac{\Pr(r=0|\bm{x}_i;\bm{\t}_0)}{\Pr(r=0|\bm{x}_i;\bm{\t}_*(\bm{\omega}))}\right\}.
$$
Rewriting $\bm{\omega}=(\bm{\omega}^\top_u,\omega_{s_*},\bm{\omega}^\top_o)^\top$ and $\bar{\mathcal{H}}_1=\{\bm{\omega}_u\in[0,1]^{s_*-1}:0<\sum^{s_*-1}_{k=1}\omega_k\leq1\}$, we further present a weighted infeasible KL loss among underfitted models as $\bar{\xi}_n=\inf_{\bm{\omega}_u\in\bar{\mathcal{H}}_1}(\sum^{s_*}_{k=1}\omega_k)^{-1}\KL(\eta_*(\bm{\omega}_u))$,
where
$$
\KL(\eta_*(\bm{\omega}_u))= \sum^n_{i=1}\left\{E(r|\bm{x}_i)\log\frac{\Pr(r=1|\bm{x}_i;\bm{\t}_0)}{\Pr(r=1|\bm{x}_i;\bm{\t}_*(\bm{\omega}_u))}+E(1-r|\bm{x}_i)\log\frac{\Pr(r=0|\bm{x}_i;\bm{\t}_0)}{\Pr(r=0|\bm{x}_i;\bm{\t}_*(\bm{\omega}_u))}\right\}.
$$
The following conditions are emerged for deriving the theorem: 
\begin{enumerate}
	\item[(A8)]~$n^{1/2}\xi^{-1}_n=o_p(S_1^{-1/2}|\mathcal{M}_{S_1}|^{-1})$.
	\item[(A9)]~The penalty factor satisfies: (i)~$\frac{n^{1/2}S^{1/2}_1|\mathcal{M}_{S_1}|}{\l_n(|\mathcal{M}_{s_*+1}|-|\mathcal{M}_{s_*}|)}=o(1)$ and (ii)~$\bar{\xi}^{-1}_n\l_n|\mathcal{M}_{S_1}|=o_p(1)$.
\end{enumerate}

Condition (A8) reflects that orders of the model size and the number of candidates are both related to KL divergence of the best-but-infeasible weighted model. It is regarded as an extension in high-dimensional case.  When $S_1$ and $|\mathcal{M}_{S_1}|$ are finite, this condition degenerates to $n\xi_n^{-2}=o_p(1)$, a classical constraint on model averaging inference (e.g., \hyperlink{ref:Ando2017}{Ando and Li 2014}; \hyperlink{ref:Zhang2016}{Zhang et al. 2016}; \hyperlink{ref:Zhang2020}{Zhang et al. 2020}; \hyperlink{ref:Fang2022}{Fang et al. 2022}). Condition (A9) ensures that the OC model can be consistently selected by SMA. Specifically, the first part in (A9) illustrates that $\l_n$ should not be a constant, while the second part shows the order is dominated by $\bar{\xi}_n$, the latter is separated from KL divergence for demonstrating the ``over-consistency'' of weights (see condition (11) in \hyperlink{ref:Chen2023}{Chen et al. 2023}).

\begin{theorem}
	Under conditions {\em (A1)--(A9)}, when $\mathcal{S}_1$ contains the OC model, we have for $n\to\infty$, {\em (i)} $\hat{\omega}_{s_*}\stackrel{p}{\to} 1$; {\em (ii)} $\hat{\bm{\t}}(\hat{\bm{\omega}})\stackrel{p}{\to}\bm{\t}_0$.
\end{theorem}

Theorem 1 guarantees the consistency of nuisance parameter estimator. Notably, this result relies on the structural condition for candidates and will raise two potential problems: (1) how to construct $\mathcal{S}_1$ to cover the OC model, and (2) whether the estimator is valid for subsequent inference when the true mechanism deviates from (\ref{sec2.eq2}). The first will be addressed in Section 5, while the second will be examined through numerical studies to assess prediction robustness.

\section{SMA for conditional quantile function}

In this section we intend to estimate (\ref{sec2.eq1}) by SMA. 
Given $0<\tau<1$, we still need to divide covariates to construct submodels. 
Denote $\mathcal{S}_2=\{\mu^\mm_i:m=1,...,S_2\}$ as the set of total $S_2$ candidates, where the $m$-th model is written as
$$
\mu^\mm_i =\bm{\b}^{\mm^\top}\bm{x}^\mm_i
= \sum_{j\in\mathcal{I}_m}\b_{j}x_{ij},
$$
and $\mathcal{I}_m\subset\{1,...,p\}$ is the index set of predictors. As mentioned previously, the SMA procedure is implemented by matching weights for candidate estimators. 
In fact, (\ref{sec2.eq2}) implies that
\begin{equation}
E\rho_\tau\left(\e\right)=
E\left\{r\rho_\tau\left(\e\right)+(1-r) M_0(\bm{x};\g)\right\},
\label{sec3.eq1}
\end{equation} 
where 
$$
 M_0(\bm{x};\g) = \frac{E_1\left\{\exp(\g y)\rho_\tau\left(\e\right)|\bm{x}\right\}}{E_1\left\{\exp(\g y)|\bm{x}\right\}},
$$
and $\rho_\tau(\e)=\e(\tau-I(\e\leq 0))$ is named as quantile loss or check loss. Detailed derivation of (\ref{sec3.eq1}) is attached in Part C of online supplementary material and, under MNAR, the parameter in $\mu^\mm_i$ hence can be consistently estimated as
\begin{equation}
\hat{\bm{\b}}^\mm_n
=\mathop{\arg\min}_{\bm{\b}^\mm\in\mathcal{B}_m} n^{-1}\sum^n_{i=1}
\left\{r_i\rho_\tau\left(y_i-\bm{\b}^{\mm^\top}\bm{x}^\mm_i\right)
+ (1-r_i)\hat{M}_0\left(\bm{x}_i;\hat{\g}(\hat{\bm{\omega}}),\bm{\b}^\mm\right)\right\},
\label{sec3.eq2}
\end{equation}
where $\hat{\g}(\hat{\bm{\omega}})$ is given in (\ref{sec2.eq4}), and $\hat{M}_0\left(\cdot\right)$ is obtained by nonparametric regression such that
$$
\hat{M}_0\left(\bm{x}_i;\hat{\g}(\hat{\bm{\omega}}),\bm{\b}^\mm\right) =
\frac{\sum^n_{j=1}r_jK_h\left(\bm{x}^\mm_j-\bm{x}^\mm_i\right)\exp(\hat{\g}(\hat{\bm{\omega}})y_j)\rho_\tau\left(y_j-\bm{\b}^{\mm^\top}\bm{x}^\mm_i\right)}
{\sum^n_{j=1}r_jK_h\left(\bm{x}^\mm_j-\bm{x}^\mm_i\right)\exp(\hat{\g}(\hat{\bm{\omega}})y_j)}.
$$ 

Let $\bm{w}=(w_1,...,w_{S_2})^\top\in\mathcal{H}_2$ and $\mathcal{H}_2=\{\bm{w}\in[0,1]^{S_2}:\sum^{S_2}_{m=1}w_k=1\}$. Similar to the procedure in Section 2.2, the SMA estimator of the $\tau$-th CQF has the form
$\sum^{S_2}_{m=1}w_m\hat{\mu}^{\mm}$. Note that the complexity of (\ref{sec3.eq2}) is almost higher than $O(pn^2)$, adopting multifolds cross-validation will sensibly deteriorate computing efficiency (e.g., the complexity of JMA is approximating $O(pn^3)$). To suppress computational costs and overfitting,
we utilize regularization to optimize $\bm{w}$, of which the criterion is modified by
$$
\widehat{\mathrm{QL}}_n(\bm{w}) =
n^{-1}\sum^n_{i=1}
\left\{r_i\rho_\tau\left(y_i-\sum^{S_2}_{m=1}w_m\hat{\mu}^\mm_i\right)
+ (1-r_i)\hat{M}_0\left(\bm{x}_i;\hat{\g}(\hat{\bm{\omega}}),\bm{w}\right)\right\}
+\frac{\p_n}{n}\sum^{S_2}_{m=1}w_m\left|\mathcal{I}_m\right|,
$$
here $\p_n$ is a penalty factor that diverges to infinity, and 
$$
\begin{aligned}
\hat{M}_0\left(\bm{x}_i;\hat{\g}(\bm{\omega}),\bm{w}\right)
& =
\frac{\hat{E}_1\left\{\exp(\g y)\rho_\tau\left(y-\sum^{S_2}_{m=1}w_m\hat{\mu}^\mm_i\right)\big|\bm{x}_i\right\}}
{\hat{E}_1\left\{\exp(\g y)|\bm{x}_i\right\}} \\
& =
\frac{\sum^n_{j=1}r_lK_h\left(\bm{x}^\mm_j-\bm{x}^\mm_i\right)\exp(\hat{\g}(\hat{\bm{\omega}})y_j)\rho_\tau\left(y_j-\sum^{S_2}_{m=1}w_m\hat{\mu}^\mm_i\right)}
{\sum^n_{j=1}r_jK_h\left(\bm{x}^\mm_j-\bm{x}^\mm_i\right)\exp(\hat{\g}(\hat{\bm{\omega}})y_j)}.
\end{aligned}
$$
Therefore, the weight is chosen by 
\begin{equation}
\hat{\bm{w}} = \mathop{\arg\min}_{\bm{w}\in\mathcal{H}_2}\widehat{\mathrm{QL}}_n\left(\bm{w}\right),
\label{sec3.eq3}
\end{equation}
and the SMA estimator of (\ref{sec2.eq1}) is
$$
\hat{Q}_\tau(y|\bm{x}) = \sum^{S_2}_{m=1}\hat{w}_m\hat{\bm{\b}}^{\mm^\top}_n\bm{x}^\mm.
$$

\section{Asymptotic behaviors}

In this section, we demonstrate the asymptotic properties of CQF estimator. We firstly establish the consistency of the candidate estimator and then study the predicting efficiency. The latter is investigated by the asymptotic performance of selected weights.	
Let $\bm{\b}^\mm_*\in\mathcal{B}_m$ be the unique minimizer of $E\rho_\tau(y-\bm{\b}^{\mm^\top}\bm{x}^\mm)$. The following conditions are required for theoretical inference:

\begin{enumerate}
	\item[(B1)]~$\mathcal{B}_m$'s are compact over $m=1,...,S_2$.	
	\item[(B2)]~$\underline{c}=\min_{1\leq m\leq S_2}\underline{c}_\mm>0$ and is nonincreasing as $n\to\infty$, where $\underline{c}_\mm$ is the minimum eigenvalue of $E\bm{x}^\mm\bm{x}^{\mm^\top}$.
	\item[(B3)]~(i) Uniformly for $\bm{x}\in\mathcal{X}$, $f_{y|x}(\cdot|\bm{x})$ is bounded away from $0$ and $\infty$ uniformly on its support; (ii) the marginal densities of $x_j$'s ($j=1,...,p$) are uniformly bounded away from $0$ and $\infty$. 
	\item[(B4)]~Let $|\bar{\mathcal{I}}|=\max_{1\leq m\leq S_2}|\mathcal{I}_m|$, (i)~$|\bar{\mathcal{I}}|^{1/2}(\hat{\g}(\hat{\bm{\omega}})-\g_0)=o_p(1)$; (ii)~$|\bar{\mathcal{I}}|^{\kappa_1}n^{-\kappa_2}=o(1)$, where $\kappa_1$, $\kappa_2$ are both positive.	
\end{enumerate}

Condition (B1) is similar to (A1), and (B2) is common in high-dimensional random matrix theory. It is easy to see that $\underline{c}$ is inversely proportional to $n$ while the order is refined to promote the consistency. Condition (B3) is sufficient for the concentration of estimators. Condition (B4) restricts the order of candidate model size.

\begin{lemma}
	Under conditions {\em (A1)--(A9)}, {\em (B1)--(B4)}, when $\mathcal{S}_1$ contains the OC model, then for any $C>0$, there exist positive constants $C_4$ and $C_6$ such that
	\begin{equation}
	\Pr\left(\max_{1\leq m\leq S_2}\left\|\hat{\bm{\b}}^\mm_n-\bm{\b}^\mm_*\right\| \geq C \left|\bar{\mathcal{I}}\right|^{\kappa_1}n^{-\kappa_2}\right)
	\leq 2S_2\exp\left(-C_4\underline{c}^2 n^{1-4\kappa_2}\left|\bar{\mathcal{I}}\right|^{4\kappa_1-2}\right)+
	S_2\exp\left(-C_6n\right).
	\label{sec4.eq1}
	\end{equation}
\end{lemma}

Above lemma shows the tail behavior of estimators uniformly over candidate models. One benefit of this result is to determine the optimal order of $\left|\bar{\mathcal{I}}\right|$ and $S_2$ separately. Note that (\ref{sec4.eq1}) does not directly reveal the consistency of parameter estimators unless further constraints are imposed. For instance, if we regard $\underline{c}=O(1)$ and $\left|\bar{\mathcal{I}}\right|=n^\a$ for $\a>0$, to promote the convergence of the exponential tail, one can get a necessary condition that $\a\in\left(0, 1/2\right)$ (detailed discussion is found in the proof of Lemma 3). Of such case considered, $\log(S_2)=o\left(n+n^{1-4\kappa_2}\left|\bar{\mathcal{I}}\right|^{4\kappa_1-2}\right)$, which means the number of candidates can be grown by an exponential order as well. 

Next, we establish theoretical properties of $\hat{\bm{w}}$. Denote the out-of-sample finial prediction risk (FPR) of weighted CQF as
$$
\mathrm{FPR}_{n,\tau}\left(\bm{w}\right) = 
E\left\{\rho_\tau\left(y-\sum^{S_2}_{m=1}w_m\hat{\bm{\b}}^{\mm^\top}_n\bm{x}^\mm\right)\bigg|\mathcal{D}_n\right\}.
$$
Following conventional frameworks of frequentist model averaging, we assume additional conditions.

\begin{enumerate}
	\item[(B5)]~$E\mu^4=O(1)$.
	\item[(B6)]~(i) $n^{-1}\left|\bar{\mathcal{I}}\right|^2=o(\underline{c}^2\left|\bar{\mathcal{I}}\right|^{4\kappa_1}n^{-4\kappa_2})$; (ii) $S_2\lesssim n^{2\kappa_2}\left|\bar{\mathcal{I}}\right|^{-2\kappa_1-1}$; (iii)	$\p_n\left|\bar{\mathcal{I}}\right|=o(n)$.
\end{enumerate}

Condition (B5) is same as assumption 4 (iii) in \hyperlink{ref:Wang2023}{Wang et al. (2023)}. Condtion (B6) (i) guarantees an exponentially decaying tails in above lemma, and (iii) theoretically prevents the regularization term from dominating the weighting process. Condition (B6) (ii) determines the maximum dimensionality of candidates to establish the asymptotic optimality. Specially, for finite-dimensional candidate models it demonstrates that $\mathcal{S}_2$ can accommodate up to $O(n^{1/2})$ estimators with exponential tail. In high-dimensional settings, the dimensionality bounds can be determined through the following lemma:

\begin{lemma}
	Suppose the size of the largest candidate of CQF has the order $n^{\a}$, if Lemma {\em 2} holds with condition {\em (B6)}, then $\a\in(0,1/4)$. 
\end{lemma}

Lemma 3 verifies the divergence of $|\bar{\mathcal{I}}|$ under complexity order assumptions, and the upper-bound is consistent with that in most high-dimensional inference. Crucially, (B6) (ii) is satisfied with $\phi_n \lesssim n^{1-\a}$, which takes an analogous penalty restriction in model selection (\hyperlink{ref:Shao1997}{Shao 1997}). 
Consequently, we establish the following property:

\begin{theorem}[Asymptotic optimality]
	Given $0<\tau<1$, conditions {\em (A1)--(A9)} and {\em (B1)--(B6)}, when $\mathcal{S}_1$ contains the OC model, {\em (\ref{sec3.eq3})} is asymptotically optimal in the sense that
	$$
	\frac{\mathrm{FPR}_{n,\tau}\left(\hat{\bm{w}}\right)}
	{\inf_{\bm{w}\in\mathcal{H}_2}\mathrm{FPR}_{n,\tau}\left(\bm{w}\right)}= 1+O_p\left(R_n+n^{-1}\p_n|\bar{\mathcal{I}}|\right)
	=1+o_p(1),
	$$
	where $R_n=\sum^5_{l=1}r_{ln}$ and $r_{ln}$'s are given in the proof.
\end{theorem}

Theorem 2 reveals the primal result in model averaging theory. When all candiates in $\mathcal{S}_2$ are misspecified, it investigates that the SMA procedure has the optimal predicting accuracy in the sense that its FPR is asymptotically identical to that of the best-but-infeasible averaging estimator. Although weights are selected from respondents, FPR is calculated with fully out-of-sample observations. The property is asymptotically equivalent to that obtained in the full case.
In fact, above theorem implies a principle that $|E\rho_\tau(\e)|$ is bounded away from $0$ and infinity, further requires $E\e\neq0$ for $\tau\in(0,1)$. To cover this situation we reform asymptotic optimality via ``excess final prediction risk'' (\rm{EFPR}) defined as
$$
\mathrm{EFPR}_{n,\tau}(\bm{w}) = \mathrm{FPR}_{n,\tau}(\bm{w})-E\rho_\tau(\e).
$$
As discussed in \hyperlink{ref:Xu2023}{Xu et al. (2023)}, $\mathrm{EFPR}_{n,\tau}(\bm{w})\geq 0$. The advantage of \rm{EFPR} is given as an approximating unbiasedness in quantile framework, while it induces an extra order of $\inf_{\bm{w}\in\mathcal{H}_2}\rm{EFPR}_{n,\tau}(\bm{w})$. Based on Theorem 2, it can be naturally promoted into our SMA estimator by the following corollary:
\begin{corollary}
Let 
$$
\zeta_n = \inf_{\bm{w}\in\mathcal{H}_2}E\left\{\int^{\sum^{S_2}_{m=1}w_m\hat{\mu}^\mm-\mu}_0\left[F_{y|\bm{x}}(s)-F_{y|\bm{x}}(0)\right]\mathrm{d}s\bigg|\mathcal{D}_n\right\}.
$$
Under same conditions in Theorem {\em 2}, if $\zeta^{-1}_n(R_n+n^{-1}\p_n|\bar{\mathcal{I}}|)=o_p(1)$, {\em (\ref{sec3.eq3})} is asymptotically optimal in the sense that
$$
\frac{\mathrm{EFPR}_{n,\tau}\left(\hat{\bm{w}}\right)}
{\inf_{\bm{w}\in\mathcal{H}_2}\mathrm{EFPR}_{n,\tau}\left(\bm{w}\right)}= 1+o_p\left(1\right).
$$
\begin{remark}
It is worthy noting that asymptotic optimalities are established by the premise of the existence of the OC candidate in $\mathcal{S}_1$. This is because {\em (\ref{sec3.eq1})} yields an unbiased estimating equation iff the response probability is correctly specified. However, above properties may persist even under model misspecification in $\mathcal{S}_1$. In such scenario, $\hat{\bm{\b}}^\mm_n$'s are no longer consistent to the pseudo-true parameters associated with the full case, and further propagates to {\em FPR}, {\em EFPR}.
\end{remark}
\end{corollary}

Hereafter we focus on the property of $\hat{\bm{w}}$ in the sense that $\mathcal{S}_2$ contains correct models. We anticipate the estimator in (\ref{sec3.eq3}) achieves the same convergency to least squares model averaging weights (see in, \hyperlink{ref:Zhang2020}{Zhang et al. 2020}). Fortunately, the equivalence emerges through an algebraic transformation between quantile loss and mean squared error, and enables linearization of underfitted components. By Taylor's expansion, we have 
$$
\begin{aligned}
& \quad E\left\{\left(\sum^{S_2}_{m=1}w_m\hat{\bm{\b}}^{\mm^\top}_n\bm{x}^\mm-\mu\right)^2\bigg|\mathcal{D}_n\right\} \\
& =
\sum_{m\in\mathcal{I}_\mathrm{cor}}w_mE\left\{\left(\mu-\bar{\mu}(\bm{\omega})\right)\left(\hat{\mu}^\mm-\mu\right)\bigg|\mathcal{D}_n\right\}
+
\sum_{m\nin\mathcal{I}_\mathrm{cor}}w_mE\left\{(\mu-\bar{\mu}(\bm{\omega}))\left(\hat{\mu}^\mm-\mu\right)\bigg|\mathcal{D}_n\right\} \\
& \triangleeq RL_{1n}\left(\bm{w}\right) + RL_{2n}\left(\bm{w}\right),
\end{aligned}
$$ 
where $\bar{\mu}(\bm{w})$ lies between $\mu$ and $\sum^{S_2}_{m=1}w_m\hat{\mu}^\mm$, and $\mathcal{I}_\mathrm{cor}\subset\{1,...,S_2\}$ is the index set of correct candidates in $\mathcal{S}_2$. Moreover, denote $\bar{\mathcal{H}}_2 =\{\bm{w}\in[0,1]^{S_2}:\sum^{S_2}_{m=1}w_m=1,~\sum_{m\in\mathcal{I}_\mathrm{cor}}w_m\neq1\}$ and
$$ 
\bar{\zeta}_n = \inf_{\bm{w}\in\bar{\mathcal{H}}_2}\left(\sum_{m\nin\mathcal{I}_\mathrm{cor}}w_m\right)^{-1}RL_{2n}\left(\bm{\omega}\right).
$$ 
Similar with (A9) (ii), we impose the following condition:
\begin{enumerate}
	\item[(B7)]~$\max\left\{\left|\bar{\mathcal{I}}\right|^{\kappa_1+1/2}n^{-\kappa_2},n^{-1}\p_n\left|\bar{\mathcal{I}}\right|,R_n\right\}=o(\bar{\zeta}_n)$.
\end{enumerate}

\begin{lemma}[Over-consistency]
	Given $0<\tau<1$, conditions {\em (A1)--(A9)} and {\em (B1)--(B7)}, SMA weight estimator is over-consistent, i.e.,
	$$
	\sum_{m\in\mathcal{I}_\mathrm{cor}}\hat{w}_m \stackrel{p}{\to} 1, ~~~\text{as}~~~n\to\infty.
	$$ 
\end{lemma}

The lemma implies that weight estimators asymptotically exclude incorrect models. However, it is not enough to investigate the consistency that assigns with the optimally correct model if $|\mathcal{I}_\mathrm{cor}|>1$. Let $s_*$ be the index of the OC model, for convenience we suppose that the first $s_*-1$ candidates in $\mathcal{S}_1$ are incorrect and the last $S_2-s_*$ are correct but $|\mathcal{I}_{s_*}|<|\mathcal{I}_{m}|$ for $m=s_*+1,...,S_2$. We have the following result:
 
\begin{theorem}[Parsimony]
	Under conditions in Lemma {\em 4}, if $\p_n$ further satisfies
	$$
	\frac{\left|\bar{\mathcal{I}}\right|^{\kappa_1+1/2}n^{1-\kappa_2}+nR_n}{\min_{m\geq s_*}\left(\left|\mathcal{I}_m\right|-\left|\mathcal{I}_{s_*}\right|\right)\p_n}=o(1),
    $$ 
    {\em (\ref{sec3.eq3})} asymptotically concentrates on the OC candidate, i.e.,
	$$
	\hat{w}_{s_*} \stackrel{p}{\to} 1, ~~~\text{as}~~~n\to\infty.
	$$ 
\end{theorem}
 
Theorem 3 presents that SMA can control the number of nonzero coefficient estimator to mitigate overfitting in high-dimensional settings, contributing it as a parsimonious model averaging approach. The divergence constraint for $\p_n$ highlights the necessity of incorporating the regularization term, which is by no means contradictory to (B6) (iii) because $|\bar{\mathcal{I}}|^{\kappa_1+1/2}n^{-\kappa_2}+R_n=o(1)$. Adopting theoretical bounds of $\p_n$ requires very large $n$ and $|\bar{\mathcal{I}}|$, it is able to empirically select the order to balance computational flexibility (e.g., $\p_n=O(\log(n))$).
\section{Implement}

In previous sections we illustrate the SMA procedure and its propertie, where the estimation of $E(\cdot|\bm{x},r=1)$ is obtained by nonparametric regression. 
The above method is advantageous for theoretically demonstrating the order of model size, however, it increases the complexity of objective functions and further leads to numerical confusions. In this section we consider an alternative semiparametric procedure, which is implemented by the idea of nonparametric sampling for conditional expectation.

Observed that for a measurable $g(\bm{x},y)$ and a large scalar $L$,
$$
E_1(g(\bm{x},y)|\bm{x}_i)\approx L^{-1}\sum^L_{l=1}g(\bm{x}_i,\hat{y}_l),
$$
where $\{\hat{y}_l:l=1,...,L\}$ is a set of random draws from the distribution of $y|\bm{x},r=1$. Based on the fact that ${f}_1(y|\bm{x})\propto f(y,\bm{x},r=1)$, the latter can be definitely estimated by kernel density technique. For a  smoothed multivariate kernel function $K_h$, we have 
\begin{equation}
\hat{f}_1(y|\bm{x})\propto\hat{f}(y,\bm{x},r=1)
= \frac{1}{nh^{p+1}}\sum^n_{j=1}r_jK_h\left(\bm{z}-\bm{z}_j\right),
\label{sec5.eq1}
\end{equation}
where $\bm{z}=(\bm{x}^\top,y)^\top$. In this sense, $(\ref{sec2.eq3})$ is reformed as
\begin{equation}
\hat{m}_1(\bm{x}_i;\g) = \log\left\{L^{-1}\sum^L_{l=1}\exp(\g\hat{y}_l)\right\},
\label{sec5.eq2}
\end{equation}
and conditional expectations in (\ref{sec3.eq2}), (\ref{sec3.eq3}) are replaced by
\begin{equation}
\begin{aligned}
& \hat{M}_0(\bm{x}_i;\g)=\frac{\sum^L_{l=1}\exp(\g\hat{y}_l)\rho_\tau\left(\hat{y}_l-\mu_i\right)}{\sum^L_{l=1}\exp(\g\hat{y}_l)}, \\
& \hat{M}_0\left(\bm{x}_i;\hat{\g}(\bm{\omega}),\bm{w}\right)
=
\frac{\sum^n_{l=1}\exp(\hat{\g}(\hat{\bm{\omega}})\hat{y}_l)\rho_\tau\left(\hat{y}_l-\sum^{S_2}_{m=1}w_m\hat{\mu}^\mm_i\right)}
{\sum^n_{l=1}\exp(\hat{\g}(\hat{\bm{\omega}})\hat{y}_l)}.
\end{aligned}
\label{sec5.eq3}
\end{equation}
The SMA procedure hence can be numerically executed through the following steps:
\begin{enumerate}[leftmargin=65pt]
\item[\bf Step 1.] Draw $\{\hat{y}_l:l=1,...,L\}$ from (\ref{sec5.eq1}).
\item[\bf Step 2.] Implement SMA procedure to estimate $\bm{\t}$, where $\hat{m}_1(\bm{x}_i;\g)$ is given in (\ref{sec5.eq2}).
\item[\bf Step 3.] Implement SMA procedure for CQF, where $\hat{M}_0(\bm{x}_i;\g)$ and $\hat{M}_0\left(\bm{x}_i;\hat{\g}(\bm{\omega}),\bm{w}\right)$ are given in (\ref{sec5.eq3}).
\end{enumerate}

In numerical studies we employ Markov Chain Monte-Carlo (MCMC) sampling in step 1.
However, the proposed methods with high-dimensional covariates inherently face the curse of dimensionality. Firstly, nonparametric components suffer from an exponentially deteriorating efficiency as dimension increases. To address this we use sufficient dimension reduction to prepare a $p\times d$ matrix $B_{y|x,r=1}$ satisfying $y\perp\bm{x}|B^\top_{y|x,r=1}\bm{x},r=1$, hence the conditional expectation in Sections 2--3 is replaced by $E_1(\cdot|B^\top_{y|x,r=1}\bm{x})$. Specifically we can apply {\ttfamily dr} package in {\ttfamily R} to determine the central subspace.

Secondly, one need to constrain the cardinality of the nested candidate set $\mathcal{S}_1$ to relieve computational burden. We implement this via feature screening, which is efficiently executed by existing programs (e.g., {\ttfamily screenIID} function in {\ttfamily R}). Since the screened index set is prone to severe overfitting, we wish to develop a shrinkage protocol to refine survived predictors. An enhanced algorithm is proposed in Appendix. 

\section{Numerical studies}\label{sec4}

In this section, we conduct numerical experiments and empirical analysis to compare the proposed estimation against existing model selection and model averaging methods. We implement MCMC sampling with $500$ total iterations and take the last $50$ draws as $\{\hat{y}_l:l=1,...,50\}$. We denote the estimator from Sections 2--3 as \rm{SMA}, and the estimator from Section 5 as \rm{SMA-s}. 
The competitive methods include: (i) AIC; (ii) BIC; (iii) smoothed AIC (denoted by SAIC); (iv) smoothed BIC (denoted by SBIC); (v) averaging estimator via quantile regression information criterion proposed by \hyperlink{ref:Lu2015}{Lu and Su (2015)} (denoted by QRIC); (vi) averaging estimator via $5$-folds cross-validation without regularization (denoted by CVMA); (vii) averaging estimator by inverse probability weighting (IPW) criterion (denoted by IPW-MNAR); (viii) averaging estimator by IPW criterion under MAR specification $\mathrm{logit}\{\pi_i(\bm{\t})\}=\bm{\t}^\top\bm{x}_i$
(denoted by IPW-MAR). For (iii) and (iv), corresponding weight estimators are assigned as
$$
\hat{w}^\mathrm{SAIC}_m=\frac{\exp(-\frac{1}{2}\mathrm{AIC}_m)}{\sum^{S_2}_{m=1}\exp(-\frac{1}{2}\mathrm{AIC}_m)},~~~~~~
\hat{w}^\mathrm{SBIC}_m=\frac{\exp(-\frac{1}{2}\mathrm{BIC}_m)}{\sum^{S_2}_{m=1}\exp(-\frac{1}{2}\mathrm{BIC}_m)}.
$$
For (v) and (vi), the dominant term in the criterion is same as that in $\widehat{\mathrm{QL}}_n(\bm{w})$, while the penalty in (vi) is adopted as in (3.9) of \hyperlink{ref:Lu2015}{Lu and Su (2015)}. For (vii), the weight is optimized as 
$$
\hat{\bm{w}}=\mathop{\arg\min}_{\bm{w}\in\mathcal{H}_2} n^{-1}\sum^n_{i=1}
\frac{r_i}{\pi\left(\bm{x}_i,y_i;\hat{\bm{\t}}_n\right)}\rho_\tau\left(y_i-\sum^{S_2}_{m=1}w_m\hat{\mu}^\mm_i\right)
+\frac{\p_n}{n}\sum^{S_2}_{m=1}w_m\left|\mathcal{I}_m\right|,
$$ 
where $\hat{\bm{\t}}_n$ is obtained by SMA and each $\hat{\mu}^\mm_i$ is obtained by IPW. For (viii), the response probability is estimated by group-lasso in R package {\ttfamily grplasso}. The kernel function is chosen by multivariate Gaussian with $h = 1.5n^{-1/3}\hat{\sigma}_z$, where $\hat{\sigma}_z$ is the estimated standard deviation of observations. Besides, penalty terms in (ii), (iv), (vii) and (viii) are equally set as $\phi_n=\log(n)$.

It should be pointed out that for SMA, SMA-s and (i)--(vi), the estimation procedure of $\g$ is same as that of QR model. For example, $\hat{Q}_\tau(y|\bm{x})$ and $\hat{\g}_n$ are both estimated by AIC in (i). To save the computational cost, all competitors are implemented with MCMC sampling estimation for $E_1(\cdot|\bm{x})$.

\subsection{High-dimensional simulation design}

We consider the following data generating process:
$$
y_i = c\sum^p_{j=1}\b_jx_{ij}+\e_i,\quad\quad i=1,...,n.
$$
Here the sample size $n$ is set as either $80$ or $150$, each of $\bm{x}_i$'s are i.i.d. from $N(\bm{0}_p,\bm{I}_p)$ and $p=5n$. The error term are generated by the following scenarios: 
\begin{enumerate}[leftmargin=0pt]
	\item[]~{\bf Scenario I (heteroskedasticity):} $\e_i=e_i\sum^p_{j=p-9}x^2_{ij}/8$, where $e_i\sim N(0,1)$;
	\item[]~{\bf Scenario II (homoscedasticity):} $\e_i=e_i$, where $e_i\sim N(0,1)$.
\end{enumerate}
The active linear components are indexed as $j=1,...,s_0$ with $s_0=p/10$, $\b_j=(-1)^j j^{-1}$ for $j=1,...,s_0$ and $\b_j=0$ for $j=s_0+1,...,p$.
As designed in many model averaging literature, $c$ is supplied to control the population $R^2=\{\mathrm{Var}(y_i)-\mathrm{Var}(e_i)\}/\mathrm{Var}(y_i)$ such that $R^2$ varies across $\{0.1,0.3,0.5,0.7,0.9\}$. 

To produce nonignorable nonresponse, we use $2$ response mechanisms: 
\begin{enumerate}[leftmargin=0pt]
\item[]~{\bf Case (a) (logistic):} $-\mathrm{logit}\{\pi_i(\bm{\t})\} = \t_1x_{1i}+\t_2x_{2i}+\t_3+\t_4y_i$; 
\item[]~{\bf Case (b) (log-log):} $\log\{-\log(1-\pi_i(\bm{\t}))\} = \t_1x_{1i}+\t_2x_{2i}+\t_3+\t_4x_{1i}\sin(y_i)$.
\end{enumerate} 
In each case, the true value of the parameter is set by: (a) $\bm{\t}=(1, -1, -1, 0.7)^\top$ in {\bf scenario I}; $\bm{\t}=(1, -1, -1, 1)^\top$ in {\bf scenario II}, and (b) $\bm{\t}=(1,-1,0.3,1)^\top$ in {\bf scenario I}; $\bm{\t}=(1,-1,0.3,0.8)^\top$ in {\bf scenario II}.
We uniformly apply the logistic specification for MNAR to evaluate $\bm{\t}$ during estimations (i)--(vii), SMA and SMA-s. It can be seen that the mechanism is correctly specified in case (a), while it is structurally misspecified in case (b). On the other hand, there is always a misspecification in (viii).
To check the robustness for case (b), we additionally design a SMA procedure based on the true response mechanism, denoted as SMA-T.
The variation of the missing rate with $R^2$ are visualized in Part A of online supplementary material, which floats around $31\%-37\%$.

To make preparations for constructing candidate models, we employ the idea of \hyperlink{ref:Zhang2020CSDA}{Zhang et al. (2020)} to screen covariates. Specifically, we implement DC-SIS method (\hyperlink{ref:Li2012}{Li et al. 2012}) to estimate $\mathcal{A}_{ry|x}\triangleeq\{j:ry\text{ functional depends on }x_j\}$, which retains the top $d_n=\lfloor2n/(3\log(n))\rfloor$ corresponding to a decreasing order of distance correlation. Set $\hat{\mathcal{A}}_{ry|x}\triangleeq\{j_1,...,j_{d_n}\}$, the linear components in the $m$-th nested models of $\mathcal{S}_1$ and $\mathcal{S}_2$ are identically formed as: $\{x_{j_1},...,x_{j_{2m}}\}$ for $1\leq m<\lceil d_n/2\rceil$ and $\{x_{j_1},...,x_{j_{d_n}}\}$ for $m=\lceil d_n/2\rceil$.
Besides, we use sliced inverse regression to compute $\hat{B}_{y|x,r=1}$.

Based on $200$ independent replications, we assess the normalized final prediction risk to quantify the performance of CQF prediction, which is given as their FPR divided by the optimal full-case single-estimator (\hyperlink{ref:Hansen2007}{Hansen 2007}, \hyperlink{ref:Lu2015}{Lu and Su 2015}). FPR under the $r$-th replication is calculated by 100 pairs of out-of-sample observations $\{(\dot{\bm{x}}^\top_{s,r},\dot{y}_{s,r})^\top:s=1,...,100\}$ and is further averaged as
$$
\mathrm{FPR} = \frac{1}{200}\sum^{200}_{r=1}\frac{1}{100}\sum^{100}_{s=1}
\rho_\tau\left(\dot{y}_{s,r}-\sum^{S_2}_{m=1}\hat{w}_{m,r}\hat{\bm{\b}}^{\mm^\top}_{n,r}\dot{\bm{x}}^\mm_{s,r}\right).
$$

\begin{figure}[H]
	\subfigure{
		\begin{minipage}[t]{0.5\linewidth}
			\centering
			\includegraphics[width=1\textwidth,height=1\textwidth]{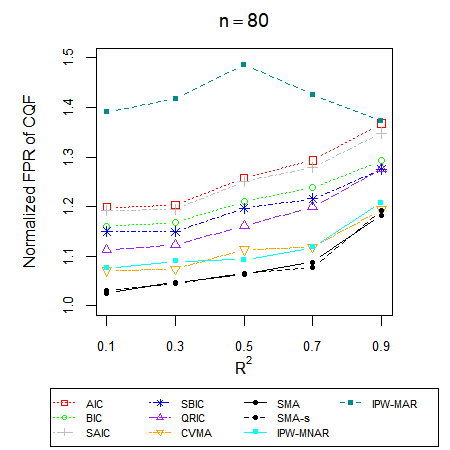}
		\end{minipage}
	}
	\subfigure{
		\begin{minipage}[t]{0.5\linewidth}
			\centering
			\includegraphics[width=1\textwidth,height=1\textwidth]{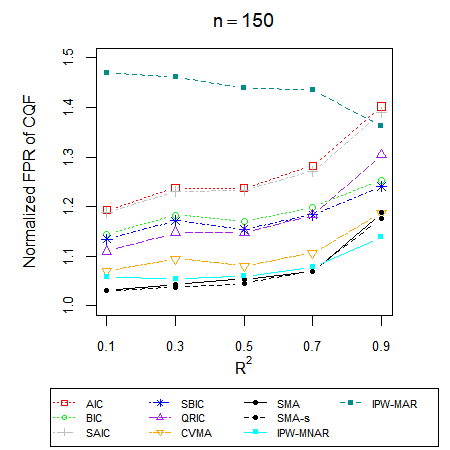}
		\end{minipage}
	}
	\caption{Normalized FPR of CQF in scenario I, case (a) with $\tau=0.05$.}\label{fig1}
\end{figure}

\begin{figure}[H]
	\subfigure{
		\begin{minipage}[t]{0.5\linewidth}
			\centering
			\includegraphics[width=1\textwidth,height=1\textwidth]{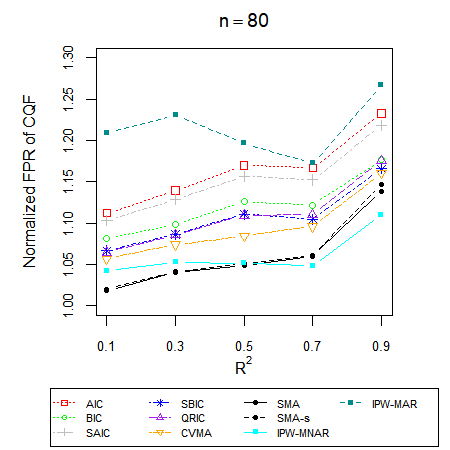}
		\end{minipage}
	}
	\subfigure{
		\begin{minipage}[t]{0.5\linewidth}
			\centering
			\includegraphics[width=1\textwidth,height=1\textwidth]{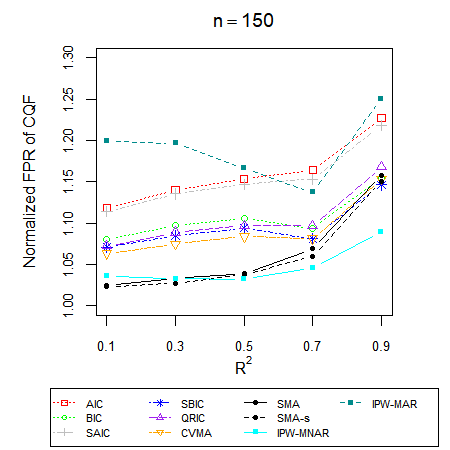}
		\end{minipage}
	}
	\caption{Normalized FPR of CQF in scenario I, case (a) with $\tau=0.5$.}
\end{figure}

\begin{figure}[H]
	\subfigure{
		\begin{minipage}[t]{0.5\linewidth}
			\centering
			\includegraphics[width=1\textwidth,height=1\textwidth]{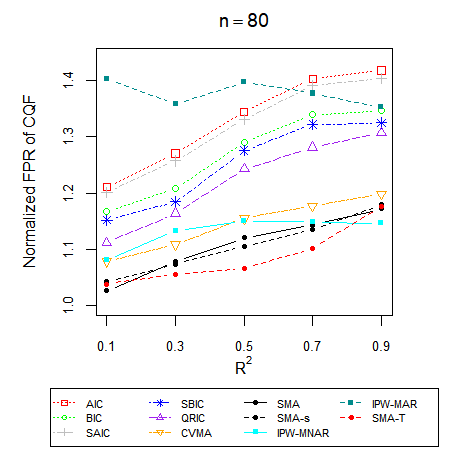}
		\end{minipage}
	}
	\subfigure{
		\begin{minipage}[t]{0.5\linewidth}
			\centering
			\includegraphics[width=1\textwidth,height=1\textwidth]{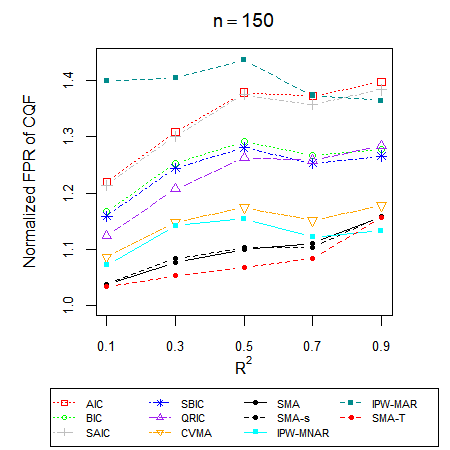}
		\end{minipage}
	}
	\caption{Normalized FPR of CQF in scenario I, case (b) with $\tau=0.05$.}
\end{figure}

\begin{figure}[H]
	\subfigure{
		\begin{minipage}[t]{0.5\linewidth}
			\centering
			\includegraphics[width=1\textwidth,height=1\textwidth]{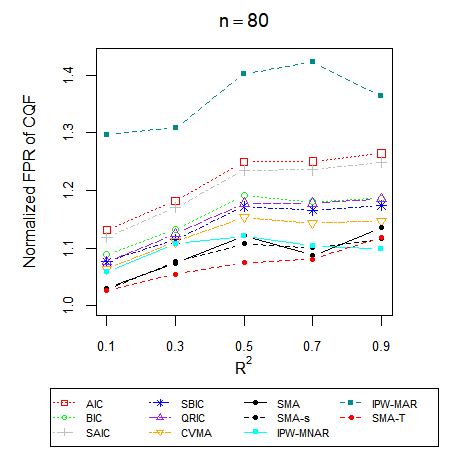}
		\end{minipage}
	}
	\subfigure{
		\begin{minipage}[t]{0.5\linewidth}
			\centering
			\includegraphics[width=1\textwidth,height=1\textwidth]{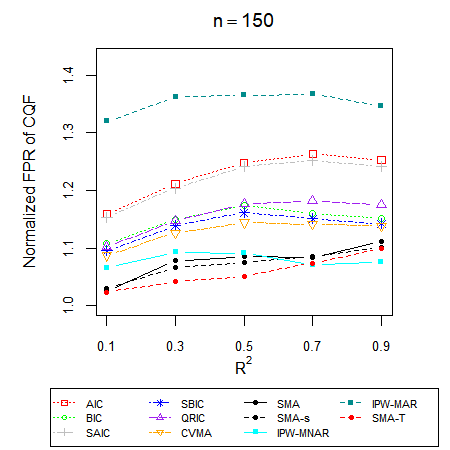}
		\end{minipage}
	}
	\caption{Normalized FPR of CQF in scenario I, case (b) with $\tau=0.5$.}\label{fig4}
\end{figure}

\begin{figure}[H]
	\subfigure{
		\begin{minipage}[t]{0.5\linewidth}
			\centering
			\includegraphics[width=1\textwidth,height=1\textwidth]{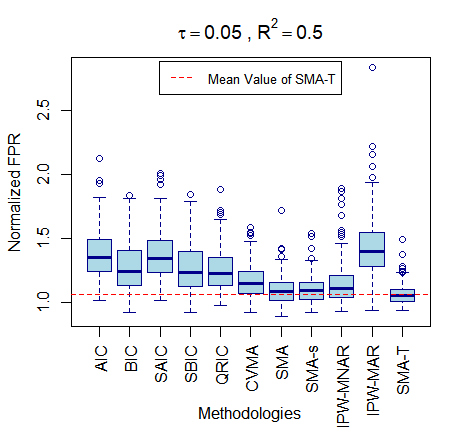}
		\end{minipage}
	}
	\subfigure{
		\begin{minipage}[t]{0.5\linewidth}
			\centering
			\includegraphics[width=1\textwidth,height=1\textwidth]{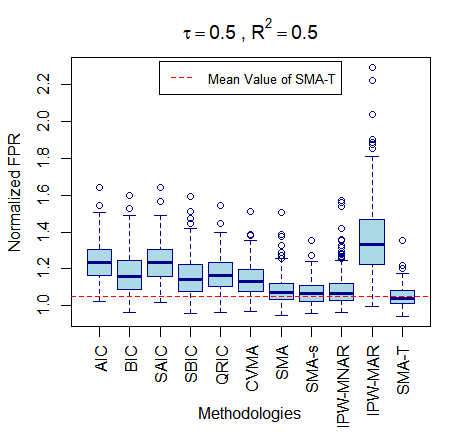}
		\end{minipage}
	}
	\caption{Boxplots of normalized FPR for CQF under heteroskedasticity and misspecified response mechanism (scenario I, case (b)), with $150$ sample size.}\label{fig5}
\end{figure}

To save space, we only report results for scenario I in Figures \ref{fig1}--\ref{fig4} (the other is found in supplementary material), which are summarized as:
(1) When $\pi_i(\bm{\t})$ is correct, two SMA estimators have outperformance than model selections, SAIC, SBIC, QRIC, CVMA and IPW-MAR, particularly in $n=80$. For $(\tau,n,R^2)=(0.05,150,0.9)$, SMAs performs close to CVMA and, for $(\tau,n,R^2)=(0.5,150,0.9)$ SMAs performs close to BIC, SBIC and CVMA.
(2) When $\pi_i(\bm{\t})$ is misspecified, two SMA estimators are significantly superior to others except IPW-MNAR, such advantage increases as $n$ decreasing. Particularly, SMA is more close to SMA-T for most of $R^2$.
(3) Compared to IPW-MNAR, SMA generally performs better for $R^2<0.9$.

Above phenomenons reflect that SMA combines parsimony of BIC with the asymptotic optimality of CVMA. The outperformance of SMA over CVMA can be attributed to two key factors: firstly, multi-folds CVMA requires a data splitting process, which may compromise numerical precision with small sample size; secondly, CVMA incurs significantly higher computational costs. Note that IPW-MAR consistently exhibits the poorest performance, indicating that the misjudgement of the type of missingness can substantially impair prediction accuracy. 
Furthermore, the comparison between SMA and IPW-MNAR reveals that the latter suffers from prediction instability, particularly when the fitted model has lower $R^2$ or the sample size is small. This instability arises from extreme estimators of propensity score.
To further illustrate the robustness under misspecified mechanism, Figure \ref{fig5} shows boxplots for the case that $(n, R^2)=(150, 0.5)$. It is clearly shown that SMA has smaller estimation variance than IPW-MNAR. This advantage arises because SMA compensates for bias induced by leveraging distributional information from the observed respondents.

\subsection{Performance of weight consistency}

In this section, we design a simulation process to illustrate the consistency of SMA weight estimators. For the sake of space, the detail discussion is presented in Part A.2 of online supplementary material.

\subsection{Empirical application}\label{sec5}

In this section we apply the SMA method for a micro-array genetic dataset. The detail discussion is presented in Part B of online supplementary material.

\section*{Conclusion}

In the background of nonignorable nonresponse, we propose a semiparametric model averaging criterion for high-dimensional quantile regression. We demonstrate that the weight estimator is parsimonious when the optimal correct model is contained in the candidate set. The asymptotic optimality in the sense of FPR and EFPR are established as well. Furthermore, we develop two nonparametric estimators for conditional expectation. Numerical results confirmed the promise of the proposed methodology.

It should be pointed out that the asymptotic optimality of quantile model averaging theory is not strictly unbiased (even for EFPR, it has been verified to be weighted-unbiased with the weight corresponding to the conditional distribution of response variable). While \hyperlink{ref:Lu2015}{Lu and Su (2015)} established a asymptotic $C_p$-type criterion for quantile regression with homoscedastic errors, the extension to heteroscedastic models remains an open research problem. Another promising direction for future studies involves applying model averaging to obtain efficient propensity score estimation across different type of response mechanisms in nonresponse analysis.

\section*{Acknowledgements}

The authors gratefully acknowledge Prof. Xinyu Zhang for his valuable suggestions on model averaging theory. We also thank the co-editor and anonymous referees for their insightful and constructive comments that significantly improved this work.

\section*{Funding}
Xiong's work is supported by National Natural Science Foundation of China (No.12401352), Postdoctoral Fellowship Program of CPSF (No.GZC20231022). Deng's work is supported by Natural Sciences and Engineering Research Council of Canada (NSERC). Wang's work is supported by National Natural Science Foundation of China (No.12271231, 12001229).

\section*{Statements and declarations}
The authors report there are no competing interests to declare.

\appendix
\section*{Appendix. A discussion for selecting nested models}

Through the discussion of asymptotic properties for SMA, an important sufficient condition is the existence of the correct candidate in $\mathcal{S}_1$, that is, all active predictors are embraced in the largest nested model. 
For sparse linear logistic regression, one can employ the regularition path method to insert significant components. However, the nonparametric form in $\Pr(r=1|\bm{x})$ deteriorates both computational burden and numerical implementation difficulty. On the other hand, applying model-free methods to screen $x_j$'s dependence on $r$ is sensitive to the failure of efficiency under MNAR, which is mainly because active parts in $y|\bm{x},r=1$ may be incorrectly selected while they are inactive in $r=1|\bm{x},y$. In this section we discuss a fast and effect algorithm to create sub-models.

Let $\mathcal{A}_{r|x,y}=\{k:r \text{~depends on~}x_k \text{~among~} y\}$ and $\mathcal{A}_{y|x}=\{k:y \text{~depends on~}x_k\}$. Despite the considerable difficulty in selecting $\mathcal{A}_{r|x,y}$, we instead identify covariates that depend on $ry$ through feature screening, denoted as $\mathcal{A}_{ry|x}$. Existing studies have demonstrated the relationship $\mathcal{A}_{r|x,y}\bigcup\mathcal{A}_{y|x}\subset\mathcal{A}_{ry|x}$ (see in, \hyperlink{ref:Zhang2020CSDA}{Zhang et al. 2020}, \hyperlink{ref:Zhao2021}{Zhao et al. 2021}), which reduces our original problem to the evaluation of $\mathcal{A}_{ry|x}$.
Different from variable screening process, it makes more sense to quantify the discrepancy between successive nested candidates through grouped variable importance (GVI). Formally, the GVI between models of size $m$ and $m-1$ is
$$
\mathrm{GVI}(m)=D_m-D_{m-1},
$$ 
where $D_m$ measures the dependence between $\bm{x}^\mm$ and response in the $m$-th model (e.g., distance correlation in DC-SIS of \hyperlink{ref:Li2012}{Li et al. 2012}, and weighted Cram\'{e}r-von Mises distance in MV-SIS of \hyperlink{ref:Cui2015}{Cui et al. 2015}). Intuitively, GVI dynamics in the nested set reflects the incremental effect of covariates: $\mathrm{GVI}(m)>0$ if newly added covariates are influential for predicting, whereas $\mathrm{GVI}(m)\leq 0$ if relatively insignificant covariates are incorporated. This phenomenon mirrors the trade-off of the goodness-of-fit by information criteria. Another contribution for estimand $\mathcal{A}_{ry|x}$ is to assess the nested candidate for QR as well. Motivated by these properties, we propose a hybrid method integrating GVI with feature screening to determine the maximum size of nested models.

We firstly summarize the naive implementation in algorithm 1, where the sample distance correlation on $\mathbb{R}^p$ is typically adopted by DC-SIS (referred as $\widehat{\mathrm{dc}}$), which can be well-calculated in (2.10) of \hyperlink{ref:Szekely2007}{S\'{z}ekely et al. (2007)}.

\begin{algorithm}[H]
	\caption{Post-selection (PS) algorithm via GVI}
	\DontPrintSemicolon
	\SetAlgoLined
	\KwIn {Covariates $\bm{x}$, response $y$, missing indicator $r$, the number of variable filter $d_n$, GVI threshold $s_n$}
	\KwOut {The selected index set $\hat{\mathcal{A}}_S$}
	Set $\bar{y}=ry$, compute the sample distance correlation (denoted as $\hat{v}_k$) between $x_k$ and $\bar{y}$ for $k=1,...,p$. \;
	Sort $\hat{v}_k$ in decreasing order, select the top $d_n$ components and denote the index set as $\hat{\mathcal{F}}_n=\{k_1,...,k_{d_n}\}$. \;
	Set $\bm{x}^{(t)}=(x_{k_1},...,x_{k_t})^\top$, compute $\widehat{\mathrm{dc}}_t$ (where $\widehat{\mathrm{dc}}_1=\hat{v}_1$). \;
	Select $\hat{\mathcal{A}}_{S}=\{k_1\}\bigcup\{k_t:\widehat{\mathrm{GVI}}(t)\geq s_n, t=2,...,d_n\}$, and denote the selected covariates as $\hat{\bm{x}}_{\hat{\mathcal{I}}_{S}}=(x_k)_{k\in\hat{\mathcal{I}}_{S}\times 1}$. \;
\end{algorithm}

As we can see, aforementioned algorithm is implemented through GVI between adjacent variables. However, it may inherently ignore variables that exhibit weak individual correlations but demonstrate strong joint correlations. Conversely, a substantial number of predictors with negligible contributions to inter-group associations may be redundantly selected. To address the limitation, we incorporate forward selection (denoted as F-step) to supplement the search for eliminated-but-active variables in $\hat{\mathcal{F}}_n\backslash\hat{\mathcal{A}}_{S}$, and use endogenous competition (denoted as E-step) to remove those with poor grouped importance out of $\hat{\mathcal{A}}_{S}$. The enhanced algorithm is formulated as a greedy method, which can be explicitly described as follows:

\begin{algorithm}[H]
	\caption{Greedy post-selection (GPS) algorithm via GVI}
	\DontPrintSemicolon
	\SetAlgoLined
	\KwIn {Covariates $\bm{x}$, response $y$, missing indicator $r$, the number of variable filter $d_n$, GVI threshold $s_n$.}
	\KwOut {The selected index set $\hat{\mathcal{A}}_S$.}
	Implement PS algorithm to generate an initial $\hat{\mathcal{A}}_{S}$. \;
	\Repeat{No changes in $\hat{\mathcal{A}}_S$}
	{
		{\rm (Implement F-step)}: \;
		For $k^{'}\in\hat{\mathcal{F}}_n\backslash\hat{\mathcal{A}}_{S}$, compute $\widehat{\mathrm{dc}}_{\hat{\mathcal{A}}_{S}\bigcup k^{'}}$ as the sample distance correlation between $\hat{\bm{x}}_{\hat{\mathcal{A}}_{S}\bigcup k^{'}}$ and $\bar{y}$. \;
		Set the buffer set $\hat{\mathcal{A}}_{B}=\hat{\mathcal{A}}_{S}\bigcup\{\hat{k}^{'}\}$, where
		$$
		\hat{k}^{'}=\mathop{\arg\max}_{k^{'}\in\hat{\mathcal{F}}_n\backslash\hat{\mathcal{A}}_{S}}
		\widehat{\mathrm{GVI}}(k^{'}),
		$$
		and $\widehat{\mathrm{GVI}}(k^{'})=\widehat{\mathrm{dc}}_{\hat{\mathcal{A}}_{S}\bigcup k^{'}}-\widehat{\mathrm{dc}}_{\hat{\mathcal{A}}_{S}}$. \;
		\eIf{$\widehat{\mathrm{GVI}}(\hat{k}^{'})>0$}
		{
			Update $\hat{\mathcal{A}}_S = \hat{\mathcal{A}}_{B}$. \;
		}
		{
			{\rm (Implement E-step)}: \;
			\Repeat{No more indexes can be removed}
			{
				For $k^{''}\in\hat{\mathcal{A}}_{S}$, select
				$\hat{k}^{''}=\mathop{\arg\min}_{k^{''}\in\hat{\mathcal{A}}_{S}}
				\widehat{\mathrm{GVI}}(k^{''})$, where $\widehat{\mathrm{GVI}}(k^{''})=\widehat{\mathrm{dc}}_{\hat{\mathcal{A}}_{S}}-\widehat{\mathrm{dc}}_{\hat{\mathcal{A}}_{S}\backslash\{k^{''}\}}$. \; 
				\If{$\widehat{\mathrm{GVI}}(\hat{k}^{''})\leq 0$}
				{
					Update $\hat{\mathcal{A}}_{S} = \hat{\mathcal{A}}_{S}\backslash\{\hat{k}^{''}\}$. \;
				}
			}
		}
	}
	For $k\in\hat{\mathcal{A}}_S$, compute the sample distance correlation $\hat{v}^S_k$ between $x_k$ and $\bar{y}$. \;
	Update the order of elements in $\hat{\mathcal{A}}_S$ by sorting $\hat{v}^S_k$ in decreasing order.
\end{algorithm}

We employ the following DGP to simulate the proposed algorithm:
$$
y_i = \bm{\b}^\top\bm{x}_i + \e_i
$$
for $i=1,...,n$, where $\b_1=\b_3=\b_5=\sqrt{8}|\tau-0.5|$, $\b_2=\b_4=1$, $\b_6=\b_7=2(\tau-0.05)$ and $\b_j=0$ with other $j$'s. $\bm{x}_i$ is drawn from $N(\bm{0}_{p},\bm{\Sigma})$ with $\Sigma=(\rho)_{p\times p}$ and $p=1000$. The error term $\e_i$ is generated from $N(0,1)$, and the response mechanism is specified as 
$$
\mathrm{logit}(\pi_i)=x_{i2}+1.5x_{i3}-x_{i4}-y_i+1.
$$ 
The estimation of CQF is implemented by SMA with different variable screening procedures. We set thresholds $d_n=\lfloor2n/(3\log(n))\rfloor$, $s_n=\min_{j=1,...,d_n}\mathrm{GVI}(j)$, and the construction of $\mathcal{S}_1$, $\mathcal{S}_2$ are same as those in Section 5.1. 

\begin{table}[H]
	\parbox{135mm}{\caption{Simulation results: MS, FP, FN, FPR and its standard deviation (in parentheses) for different methods.}\label{tabA1}}
	\smallskip
	\centering\small
	\resizebox{135mm}{!}{
		\begin{threeparttable}
			\setlength{\tabcolsep}{16pt}
			\begin{tabular}{ccccccc}
				\toprule
				$(\tau, \rho)$ & Missingness & Method & MS\tnote{1} & $\mathrm{FP}$\tnote{2} & $\mathrm{FN}$\tnote{3} & FPR(SD)~{\scriptsize ($\times 10^2$)} \\
				\midrule
				\specialrule{0em}{2pt}{2pt}
				\multirow{3}{*}{(0.05, 0)} & \multirow{3}{*}{37.87$\%$}
				& SMA     & 14    & 9.860 & 0.860 & 78.84(18.47) \\	
				&& PS-SMA  & 8.215 & 4.475 & 1.260 & 78.97(19.36) \\
				&& GPS-SMA & 10.210 & 6.110 & 0.900 & 76.86(19.43) \\
				\specialrule{0em}{2pt}{2pt}
				\midrule
				\specialrule{0em}{2pt}{2pt}
				\multirow{3}{*}{(0.05, 0.5)} & \multirow{3}{*}{39.82$\%$}
				& SMA     & 14    & 9.010 & 0.010 & 71.70(22.42) \\	
				&& PS-SMA  & 7.045 & 2.130 & 0.100 & 61.86(18.31) \\
				&& GPS-SMA & 5.200 & 0.445 & 0.245 & 59.71(18.16) \\
				\specialrule{0em}{2pt}{2pt}
				\midrule
				\specialrule{0em}{2pt}{2pt}
				\multirow{3}{*}{(0.5, 0)} & \multirow{3}{*}{38.54$\%$}
				& SMA     & 14    & 10.43 & 1.430 & 64.00(10.40) \\	
				&& PS-SMA  & 7.860 & 4.665 & 1.805 & 62.72(10.88) \\
				&& GPS-SMA & 10.385 & 6.870 & 1.485 & 62.45(10.94) \\
				\specialrule{0em}{2pt}{2pt}
				\midrule
				\specialrule{0em}{2pt}{2pt}
				\multirow{3}{*}{(0.5, 0.5)} & \multirow{3}{*}{37.67$\%$}
				& SMA     & 14    & 9.270 & 0.270 & 58.33(11.33) \\	
				&& PS-SMA  & 6.250 & 2.250 & 1.015 & 55.72(10.79) \\
				&& GPS-SMA & 6.345 & 2.265 & 0.920 & 54.17(10.89) \\		
				\bottomrule
			\end{tabular}
			\begin{tablenotes}
				\item[\tnote{1}] MS: the number of linear predictors in the largest nested candidate.
				\item[\tnote{2}] FP (false positive): the number of insignificant variables mistakenly selected as truly active.
				\item[\tnote{3}] FN (false negative): the number of truly significant predictors erroneously classified as insignificant.
			\end{tablenotes}
		\end{threeparttable}
	}
\end{table}

We consider (a) SMA by feature-screening (denoted as SMA) and (b) SMA by algorithm 1 (denoted as PS-SMA) as competitors, carry the simulation in terms of $\rho=\{0,0.5\}$ and $\tau=\{0.05,0.5\}$, respectively, and present the results averaged through $200$ replications. From Table \ref{tabA1} we can find that all scenarios exhibit nonresponse rates exceeding 35$\%$. SMA was implemented with a fixed size $|\hat{\mathcal{F}}_n|=14$ and PS-SMA, GPS-SMA both selected predictors from $\hat{\mathcal{F}}_n$. When covariates are uncorrelated, GPS-SMA conducted a very close FN to SMA's, but PS-SMA often left out some significant parts. Conversely, when covariates are moderately correlated, GPS-SMA performed a very similar FP with PS-SMA but a smaller FN under $\tau=0.5$. In all cases GPS-SMA has a smallest FPR than others. These results imply that our greedy algorithm can efficiently enhance the post-selection of candidates, further improve the prediction of model averaging. Although the FN of GPS-SMA is unsatisfactory under $\tau=0.05$ and $\rho=0.5$, it did not affect the prediction of CQF. This mainly arises from the over-exclusion of significant variables under high missingness, however, the effect of excluded parts and response is inherently captured by other strongly correlated participants.


\end{document}